\documentclass[reprint,amsmath,amssymb,aps,superscriptaddress,prl,]{revtex4-1}

\usepackage{graphicx}
\usepackage{dcolumn}
\usepackage{bm}

\usepackage[version=3]{mhchem}

\usepackage{bm}
\usepackage[usenames, dvipsnames]{xcolor}
\usepackage{threeparttable}
\usepackage{longtable}
\usepackage{afterpage}
\usepackage{multirow}
\usepackage{color}
\AtBeginDocument{\usepackage{booktabs}}
\usepackage{url}
\usepackage[section]{placeins}
\usepackage[colorlinks=true,linkcolor=blue,citecolor=magenta,urlcolor=red]{hyperref}

\definecolor{greenGM}{rgb}{0.4353, 1, 0}
\definecolor{greenmax1}{rgb}{0.7843,0,0.7843}
\definecolor{greenmax2}{rgb}{0,0.6627,0.4118}
\definecolor{greenenv}{rgb}{0,0.8,0}
\definecolor{lightgray}{rgb}{0.65,0.65,0.65}
\definecolor{lightred}{rgb}{1,0.65,0.65}
\definecolor{red2000}{rgb}{1,0,0.2}
\definecolor{red200}{rgb}{1,0,0.4}
\definecolor{red20}{rgb}{1,0,0.6}
\definecolor{lightblue}{rgb}{0., 0.46, 0.8}
\definecolor{grayL}{rgb}{0.3,0.3,0.3}
\definecolor{grayLmk}{rgb}{0.6,0.6,0.6}
\definecolor{redL}{rgb}{1,0,0.3}
\definecolor{redLmk}{rgb}{1,0,0.6}
\definecolor{boxa}{rgb}{1,0.5,0}
\definecolor{boxs}{rgb}{0.5,0.5,0.5}
\definecolor{boxm}{rgb}{0,0.75,0.2}

\usepackage{varioref}

\begin{document}

\preprint{APS/123-QED}

\setlength{\abovecaptionskip}{0pt}

\title{Machine learning electron correlation in a disordered medium}

\author{Jianhua Ma}
\email{jm9yq@virginia.edu}
\affiliation{Charles L. Brown Department of Electrical and Computer Engineering, University of Virginia, Charlottesville, Virginia 22904 USA}

\author{Puhan Zhang}
\affiliation{Department of Physics, University of Virginia, Charlottesville, Virginia 22904, USA.}

\author{Yaohua Tan}
\affiliation{Charles L. Brown Department of Electrical and Computer Engineering, University of Virginia, Charlottesville, Virginia 22904 USA}

\author{Avik W. Ghosh}
\email{ag7rq@virginia.edu}
\affiliation{Charles L. Brown Department of Electrical and Computer Engineering, University of Virginia, Charlottesville, Virginia 22904 USA}
\affiliation{Department of Physics, University of Virginia, Charlottesville, Virginia 22904, USA.}

\author{Gia-Wei Chern}
\email{gc6u@virginia.edu}
\affiliation{Department of Physics, University of Virginia, Charlottesville, Virginia 22904, USA.}

\date{\today}

\begin{abstract}
Learning from data has led to a paradigm shift in computational materials science. In particular, it has been shown that neural networks can learn the potential energy surface and interatomic forces through examples, thus bypassing the computationally expensive density functional theory  calculations. Combining many-body techniques with a deep learning approach, we demonstrate that a fully-connected neural network is able to learn the complex collective behavior of electrons in strongly correlated systems. Specifically, we consider the Anderson-Hubbard (AH) model, which is a canonical system for studying the interplay between electron correlation and strong localization. The ground states of the AH model on a square lattice are obtained using the real-space Gutzwiller method. The obtained solutions are used to train a multi-task multi-layer neural network, which subsequently can accurately predict quantities such as the local probability of double occupation and the quasiparticle weight, given the disorder potential in the neighborhood as the input. 
\end{abstract}


\maketitle


Machine learning (ML)~\cite{mitchell97,scholkopf02,nielsen15} is  one of today's most rapidly growing interdisciplinary fields. The deep-learning neural network (NN) provides a powerful universal method for finding patterns and regularities in high-dimensional data~\cite{duda01,bishop06}. It has found successful applications in a wide variety of fields. In condensed-matter physics and materials science, notable applications include using ML to guide materials design~\cite{sumpter96,kalinin15,balachandran18} and for identification and classification of crystalline structures~\cite{reinhart17,dietz17,lubbers17,cubuk15,schoenholz16}.
Recently, ML techniques have also been taken up by researchers in the area of strongly correlated systems. The majority of such activities focus on using ML to identify phases and phase transitions in many-body systems ranging from classical statistical models~\cite{wang16,carrasquilla17,wetzel17,hu17} and quantum fermionic Hamiltonians~\cite{chng17,broecker17} to topological phases~\cite{zhang17} and many-body localization~\cite{schindler17}. In these studies, a deep-learning NN, trained with data from classical or quantum Monte Carlo simulations, is shown to be able to correctly distinguish phases and predict phase diagrams. ML trained NNs can also represent thermodynamic phases in equilibrium (Boltzmann machines)~\cite{torlai16}, or ground-state wavefunctions of quantum many-body systems~\cite{carleo17,deng17}.



In this paper, we demonstrate another application of ML in correlated electron systems, namely using NN as an efficient emulator for many-body problem solvers. Specifically, our goal is to investigate whether deep-learning NN can be trained to predict electron correlation, such as the probability of double-occupation, in a disordered medium. Our approach here is similar in spirit to those adopted in quantum chemistry and materials science communities, where the ML trained NN is used to bypass the time-consuming density functional theory (DFT) calculations~\cite{brockherde17,snyder12,snyder13,li16,yao16,li16b,schutt17}. Such activities have led to the fast prediction of molecular atomization energies~\cite{rupp12,hansen13} and efficient parametrization of interatomic force fields~\cite{behler07,bartok10,li15,botu15,chmiela17}, to name a few. 
We note in passing that similar ideas of bypassing expensive numerical calculations with ML model have also been explored in correlated electron systems, such as using ML to replace the impurity solver for DMFT~\cite{arsenault14}, or to speed up total energy calculation in Monte Carlo simulations~\cite{liu17,xu17,huang17}.

{\em Model and Method}.
We consider the disordered Hubbard model in two dimensions:
\begin{eqnarray}
	\label{eq:AH}
	\mathcal{H} = -t \sum_{ ij , \sigma} \hat{c}^\dagger_{i, \sigma} \hat{c}^{\;}_{j, \sigma}  
	+ \sum_{i, \sigma} \epsilon_i \hat{n}^{\;}_{i, \sigma}+ U \sum_i \hat{n}_{i, \uparrow} \hat{n}_{i, \downarrow}
\end{eqnarray}
where $\hat{c}^\dagger_{i,\sigma}$ is the electron creation operator with spin $\sigma =\uparrow,\downarrow$ at site-$i$, and  $\hat{n}_{i, \sigma} \equiv \hat{c}^\dagger_{i, \sigma} \hat{c}^{\;}_{i, \sigma}$ is the corresponding number operator . The first-term describes nearest-neighbor hopping of electrons. The second term denotes the random local potential. The last term is the on-site Hubbard repulsion. As in the standard Anderson model, here the site energy $\epsilon_i$ is a random number drawn uniformly from the interval $[-W/2, +W/2]$.  We work at half-filling on an $L \times L$ square lattice with periodic boundary conditions. 
The Hamiltonian Eq.~(\ref{eq:AH}), also known as the Anderson-Hubbard (AH) model, is considered a paradigmatic model for studying the interplay between strong electron correlation and disorder. 

The AH model has been intensively studied by several numerical methods, including Hatree-Fock calculations~\cite{heidarian04,shinaoka09}, quantum Monte Carlo simulations~\cite{ulmke95,ulmke97,pezzoli09,pezzoli10}, and extended dynamical mean field theory (DMFT)~\cite{dobrosavljevic97,dobrosavljevic03,byczuk05,aguiar09}. In particular, intrinsic metal-insulator transition {\em without} magnetic order can be quantitatively calculated within the DMFT framework~\cite{georges96}. For application to disordered systems, DMFT can be readily combined with the typical medium theory (TMT) in which a geometrically averaged local density of states is used to construct the electron bath~\cite{dobrosavljevic03b}. The non-magnetic phase diagram of AH model obtained from the TMT-DMFT method includes three distinct phases: a correlated metallic phase, a Mott insulating phase, and an Anderson insulating phase~\cite{byczuk05,aguiar09}. 
Importantly, the two insulating phases of the AH model have very different characters. The Mott insulator results from the strong correlation effect which prohibits electrons from hopping to the neighboring sites. On the other hand, strong disorder weakens the constructive interference that allows an electron wave packet to propagate coherently in a periodic potential, leading to the Anderson insulator. TMT-DMFT calculation shows that these two insulating phases are continuously connected~\cite{byczuk05,aguiar09}.

Real-space approaches such as variational Monte Carlo (VMC) simulations~\cite{pezzoli09,pezzoli10}, statistical DMFT~\cite{dobrosavljevic97,tanaskovic03}, and the Gutzwiller methods~\cite{andrade09,andrade10} can better cope with the crucial spatial fluctuations in low dimensional systems. Applying VMC to the 2D AH model finds a continuous transition that separates the Mott insulator from the Anderson insulator in the non-magnetic phase diagram~\cite{pezzoli09,pezzoli10}. It is worth noting that there is no sharp distinction between correlated metal and Anderson insulator in 2D. Interestingly, detailed large-scale simulations of the 2D AH model within the Brinkman-Rice formalism, where the efficient Gutzwiller method can be applied, showed that strong spatial inhomogeneity gives rise to an electronic Griffiths phase that precedes the metal-insulator transition~\cite{andrade09}.

Here we employ the Gutzwiller method to solve the AH model on a square lattice. In its original formulation, a variational wavefunction $|\Psi_G \rangle = \mathcal{P}_G |\Psi_0\rangle$ is constructed by applying a real-space projector $\mathcal{P}_G = \prod_i \mathcal{P}_i$ on the Slater determinant $|\Psi_0\rangle$ obtained from the non-interacting electron Hamiltonian~\cite{gutzwiller63}. Optimization of $|\Psi_G \rangle$ can be efficiently carried out with the so-called Gutzwiller approximation (GA)~\cite{gutzwiller63}, which becomes exact in the infinite dimension limit. 
Moreover, GA corresponds to the zero-temperature saddle point solution of the slave-boson (SB) method~\cite{kotliar86}. Indeed, by factoring out the occupation probability $P^0_i$ of the uncorrelated state, the local projector can be expressed as $\mathcal{P}_i \equiv \sum_{\alpha, \beta} \Phi_{i, \alpha\beta} /(P^0_{i, \beta})^{-1/2} |\alpha\rangle \langle \beta|$, where $\alpha$, $\beta$ are the local many-electron state, and the elements of the variational matrix $\Phi_i$ correspond to the SB coherent-state amplitude~\cite{lanata12,lanata15}. For single-band Hubbard model, $\Phi_i$ is a diagonal matrix of dimension 4, i.e. $ \Phi_i = {\rm diag}(e_i, p_{i, \uparrow}, p_{i, \downarrow}, d_i)$, and the square of these diagonal elements corresponds to the probability of empty, single (with spin $\sigma = \uparrow, \downarrow$), and double-occupied states, respectively. In the following, we consider the non-magnetic solutions of the AH model and assume $p_{i, \uparrow} = p_{i, \downarrow} = p_i$.

The GA solution for the AH model in Eq.~(\ref{eq:AH}) is obtained by minimizing the following energy functional:
\begin{eqnarray}
	\mathcal{E}(\rho_{ij}, \Phi_i) = -2 t \sum_{\langle ij \rangle} \mathcal{R}_i\,\mathcal{R}_j \, \rho_{ij} + 2 \sum_i \epsilon_i  \,\rho_{ii}, \nonumber \\
	+ U \sum_i d_i^2 + 2 \sum_i \mu_i (\rho_{ii} - p_i^2 - d_i^2).
\end{eqnarray}
Here the prefactor 2 accounts for the spin degeneracy, $\rho_{ij} = \langle \Psi_0 | c^\dagger_j c^{\;}_i | \Psi_0 \rangle$ is the single-particle density matrix, $\mathcal{R}_i = (e_i p_i + p_i d_i)/\sqrt{n_i (1-n_i)}$ is the Gutzwiller renormalization factor~\cite{gutzwiller63}, $n_i = n_{i, \uparrow} = n_{i, \downarrow}$ is the local electron density, and $\mu_i$ is the Lagrangian multiplier that enforces the Gutzwiller constraint $n_i = p_i^2 + d_i^2 = \rho_{ii}$~\cite{lanata12,lanata15}. The optimization of the density matrix, or equivalently of the wavefunction $|\Psi_0\rangle$, amounts to solving the following renormalized tight-binding Hamiltonian:
\begin{eqnarray}
	\label{eq:H_eff}
	\hat{\mathcal{H}}^* = - t \sum_{\langle ij \rangle} \mathcal{R}_i \mathcal{R}_j \hat{c}^\dagger_i \hat{c}^{\;}_j +  \sum_i (\varepsilon_i + \mu_i) \hat{n}_i.
\end{eqnarray}
The minimization with respect to SB amplitudes $\partial \mathcal{E} / \partial \Phi_i = 0$, subject to constraint $e_i^2 + 2 p_i^2 + d_i^2 = 1$, can be recast into an eigenvalue problem for each site. These two steps, optimization of $\Psi_0$ and $\Phi_i$, have to be iterated until convergence is reached. 

\begin{figure}[b]
\includegraphics[width=0.99\columnwidth]{{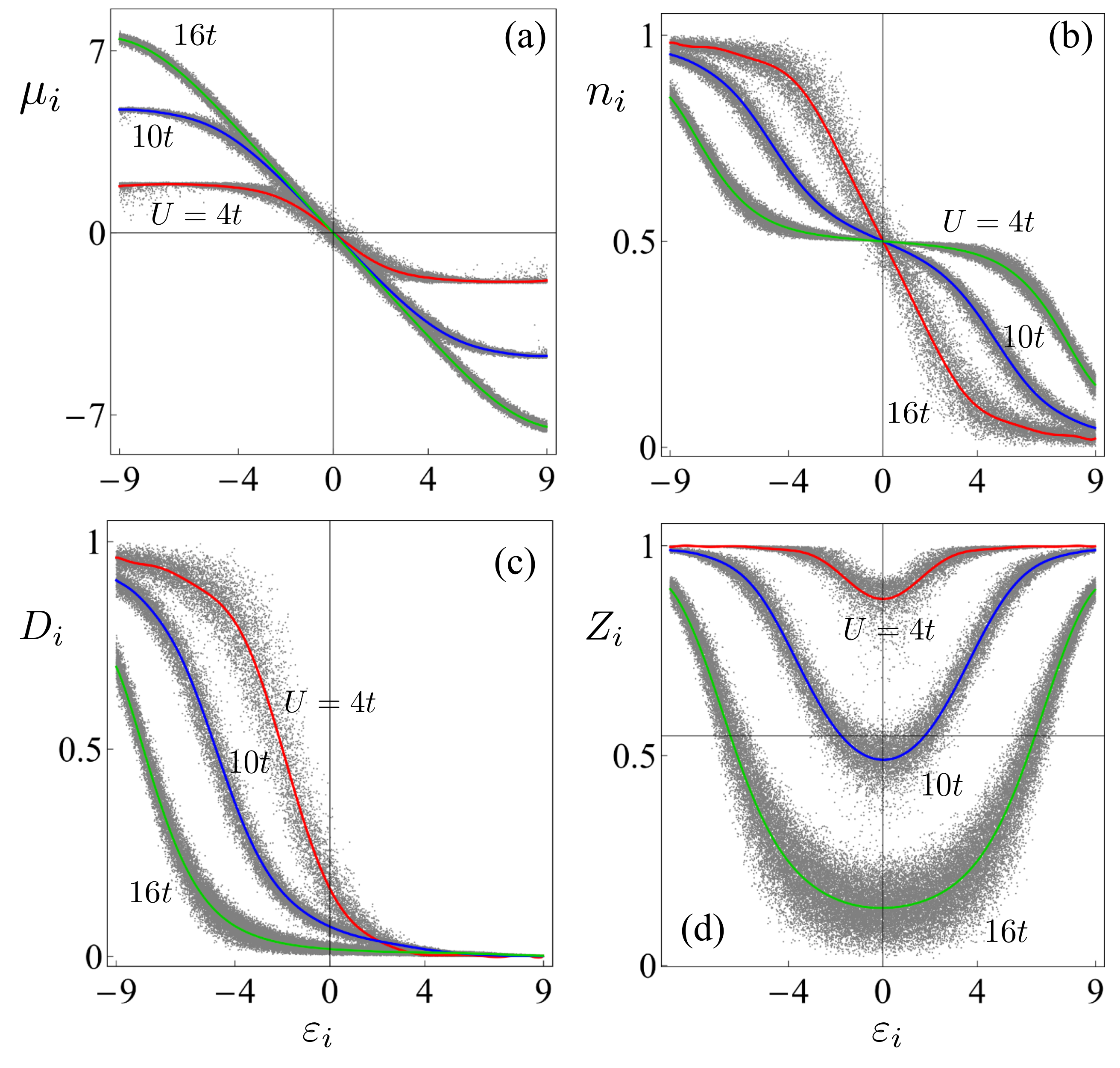}}
\caption{Summary of the GA solution for the AH model on a $30\times 30$ square lattice. The panels show the scatter diagram of (a) local energy correction $\mu_i$, (b) site electron density $n_i$, (c) probability of double occupation $D_i = d_i^2$, and (d) local quasi-particle weight $Z_i = \mathcal{R}_i^2$ versus the random site energy $\varepsilon_i$. The data points were obtained from calculations with random strength $W/t = 6, 10, 14, 18$ and three different $U = 4t$, $10t$, and $16t$. The smooths curves showing the underlying overall trend for a given $U$ were obtained using polynomial regression with up to 14th-order polynomials. The red, blue, and green curves correspond to $U = 4t$, $10t$, and $16t$, respectively. 
}
\centering
\label{fig:ga-data}
\end{figure}

Using the above GA solver on a $L = 30$ square lattice, large datasets were generated with various disorder strengths $W/t = 6, 10, 14, 18$ and Hubbard parameters $U/t = 2, 4, \cdots, 16$. The scatter plots in Fig.~\ref{fig:ga-data} show the various local quantities versus the random site energy $\varepsilon_i$ obtained from the GA solution with three different values of Hubbard repulsion. The local quantities are the site Lagrangian multiplier $\mu_i$, the local electron density $n_i$, the double occupation probability $D_i = d_i^2$, and the local quasi-particle weight $Z_i = \mathcal{R}_i^2$. Interestingly, for a given $U$, the data points cluster around a smooth curve, indicating an underlying continuous trend. More quantitatively, we used polynomial regression to determine the overall dependence of the local quantities on the site energy $\varepsilon_i$; see the solid curves in Fig.~\ref{fig:ga-data}.

Extensive studies on the statistics of electron correlation in 2D AH model have been carried out using SB or real-space DMFT methods~\cite{tanaskovic03,andrade09,andrade10}. One interesting phenomenon is the screening of the impurity potential due to electron correlations, especially close to the metal-insulator transition. Our result shown in Fig.~\ref{fig:ga-data}(a) clearly demonstrates this trend. Indeed, from Eq.~(\ref{eq:H_eff}), one can define a renormalized site potential as $\tilde \varepsilon_i = \varepsilon_i + \mu_i$. The anti-correlation between $\mu_i$ and $\varepsilon_i$   thus results in a reduced effective site potential. Moreover, the local density $n_i$ exhibits a more homogeneous distribution in the vicinity of Fermi energy with increasing~$U$; see Fig.~\ref{fig:ga-data}(b). 

The overall behavior of local quasi-particle weight versus $\varepsilon$ is consistent with the result obtained from TMT-DMFT using SB method as the impurity solver~\cite{aguiar09}. As shown in Fig.~\ref{fig:ga-data}(d), electrons at large $|\varepsilon_i|$ get less renormalization, i.e. retain a larger $Z_i$, compared with those close to the Fermi energy ($\varepsilon \sim 0$). Moreover, the difference between large and small $Z_i$ increases as one approaches the Mott transition boundary. This behavior also indicates a strong spatial inhomogeneity. While electrons in some regions become localized magnetic moments characterized by a vanishing $Z_i$, electrons in other regions undergo Anderson localization transition and maintains a large value of $Z_i$. 

\begin{figure}[t]
\includegraphics[width=0.99\columnwidth]{{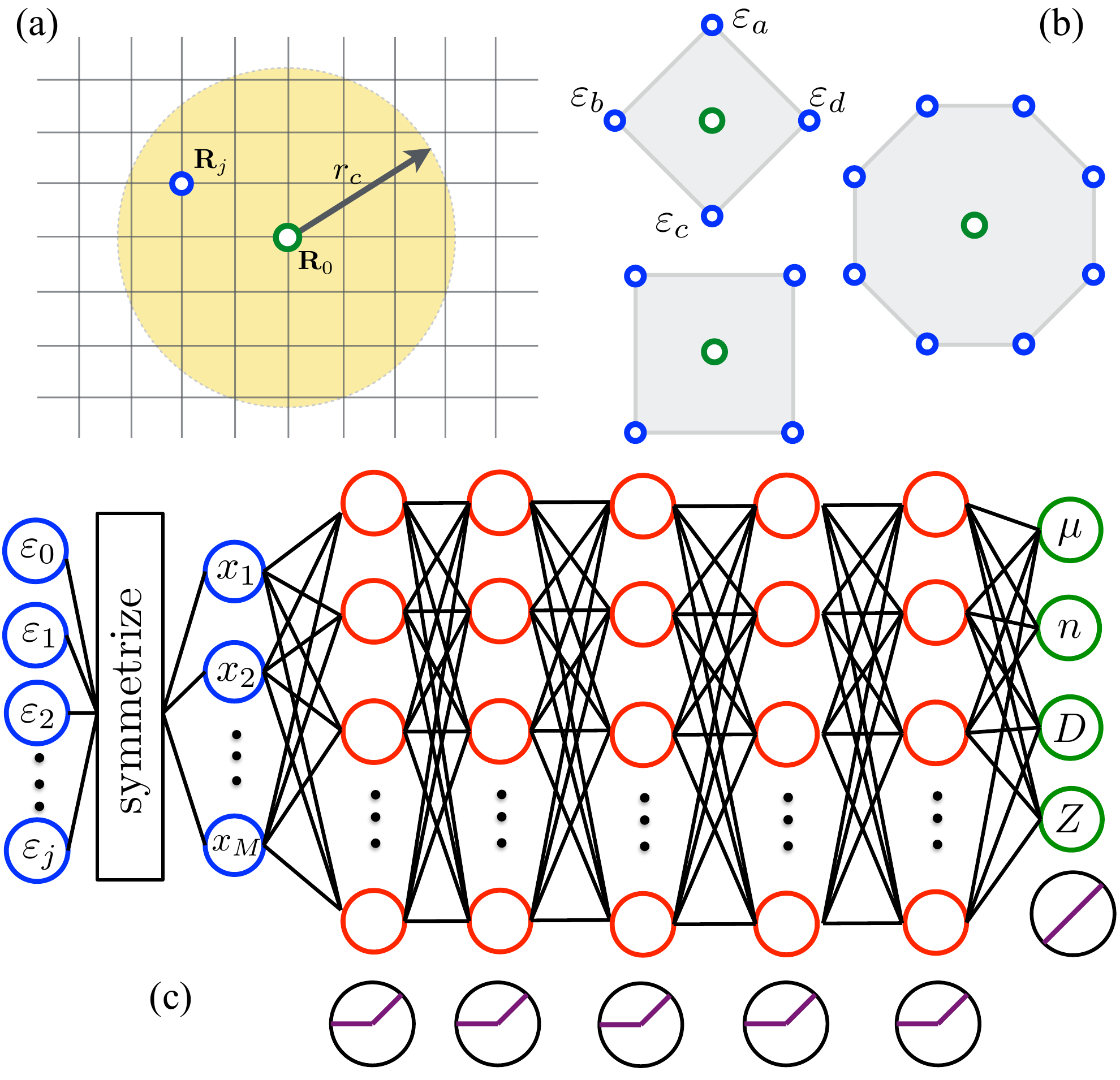}}
\caption{(a) Schematic showing the target site at $\mathbf R_0$ on the square lattice. Random potentials $\varepsilon_j$ of neighbors up to some cutoff radius $r_c$ are used as input to the neural network (NN). (b) Basic invariant subgroups of neighbors include two types of squares and a octagons. (c) Architecture of the fully-connected NN for the disordered correlated systems. For input, we use all the random-distributed on-site energies in the certain circle. ReLU activation function is used in the 5 hidden-feature extraction layers with $512\times256\times256\times128\times64$ nodes. The linear activation function is used to predict local quantities including $\mu$, $n$, $D$, and $Z$.}
\centering
\label{fig:NN}
\end{figure}

In order to capture the spatial site-to-site fluctuations of electron correlation, we next employ deep-learning techniques to predict the local electronic properties of the AH model. More specifically, our goal is to predict local quantities $\mu$, $n$, $D$, and $Z$ at a randomly picked site, say site-0, with the site potentials $\varepsilon_j$ in its neighborhood within a cutoff radius $r_c$ as the input; see Fig.~\ref{fig:NN}(a). This, of course, is based on the assumption of locality which implies that correlation functions decay strongly with the distance. In general, the single-particle density matrix exhibits an exponential and a power decay for insulators and metals, respectively. The localization of electron wavefunctions due to disorder also enhances the decay of correlation functions, especially in 2D. To quantify this locality approximation, we have repeated our ML training with various $r_c$, and have verified that the predictions of the NNs are not sensitive to the cutoff radius. The results presented below were obtained by including up to 14th nearest neighbors with a total of 89 sites within the cutoff.

A proper representation of the site energies $\varepsilon_j$ is crucial in order to provide a description of the neighborhood that is invariant under fundamental transformations of the lattice symmetry. To this end, we first decompose all $\varepsilon_j$ into irreducible representations (irrep) of point group~$D_4$, which is the site symmetry group of a square lattice. The neighboring sites can be classified into three different invariant sub-sets, as shown in Fig.~\ref{fig:NN}(b). Decomposition of these sub-sets into the corresponding irreps is straightforward. Taking the square as an example, there are three irreps: $x_{A_1} = \varepsilon_a + \varepsilon_b + \varepsilon_c + \varepsilon_d$, $x_{B_1} = \varepsilon_a - \varepsilon_b + \varepsilon_c - \varepsilon_d$, and $\mathbf x_E = (\varepsilon_a - \varepsilon_c, \varepsilon_b - \varepsilon_d)$.  The amplitudes of each irrep and their relative phases are then used as the input for the NN. For example, consider all doublet irreps: $\mathbf x_m$ with $m = 1, 2, \cdots, M$, where $M$ is the total number. The amplitudes $|\mathbf x_m|$, and relative angle $\cos\theta_{mn} = \mathbf x_m \cdot \mathbf x_n / |\mathbf x_m|\,|\mathbf x_n|$ are invariant under symmetry operations. We note that this descriptor of the site environment is similar to the atom-centered symmetry functions used in ML potentials for quantum molecular dynamics simulations~\cite{behler07,botu15}.

We design a fully-connected neural network (NN) with 5 hidden layers consisting of $n=512\times256\times256\times128\times64$  rectified linear units (ReLU) neurons~\cite{nair10}. The input layer is the symmetrized neighborhood $\varepsilon$ as discussed above. The NN performs a sequence of transformations on the input that are illustrated in Fig.~\ref{fig:NN}(c).  In the $m$-th layer, the $n$-th neuron processes the activation $\mathbf{a}^{(m-1)}$ from $(m-1)$-th layer through independent weights and biases $\mathbf{w}^{(m-1)}\mathbf{a}^{(m-1)}+\mathbf{b}^{(m-1)}$. After the ReLU functions, the outcome is fed forward to be processed by the output neuron with linear activation function. Importantly, here we adopt the multi-task ML technique~\cite{caruana97} that forces the NN to learn multiple local electron properties simultaneously. The additional constraints coming from the multi-task setup helps the search for the true ML model because of the smaller set of models that can fit all properties simultaneously.

\begin{figure}[t]
\includegraphics[width=\columnwidth]{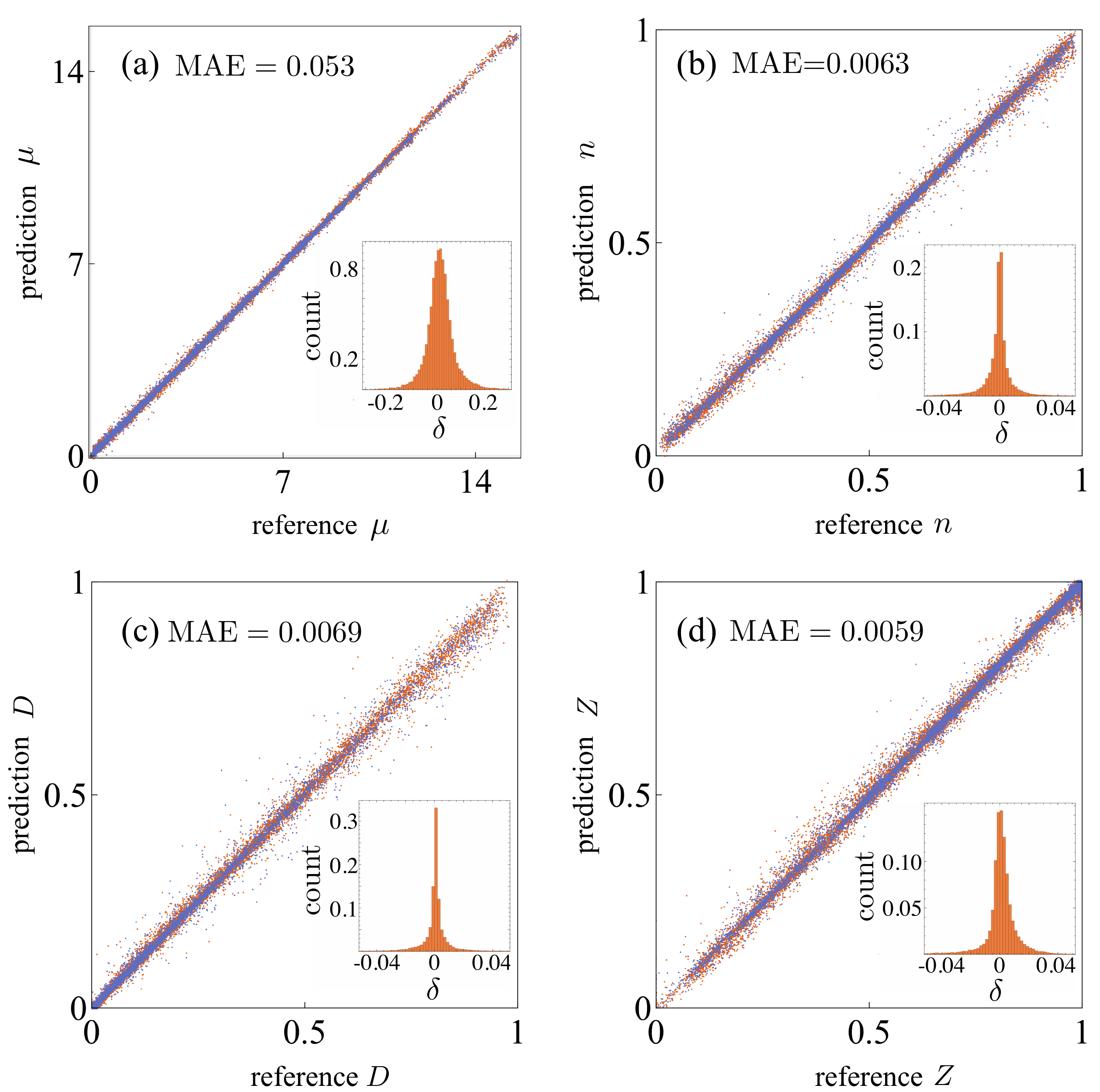}
\caption{Comparison of the ML predictions with references obtained from the GA solvers, for (a) the local potential renormalization $\mu_i$, (b) site electron density $n_i$, (c) probability of double occupation $D_i$, and (d) local quasi-particle weight $Z_i$. 
The blue and orange data points denote predictions for training and test datasets, respectively. The insets show the normalized count of the error $\delta$ defined as the difference between prediction and reference values.  
}
\centering
\label{fig:ml-prediction}
\end{figure}

We use mean absolute error (MAE) as the cost function with the L2 regularization~\cite{ng04} to avoid overfitting and a minimum batch size of 100. We use randomly mixed 900000 data samples as the training set and perform a 5-fold cross-validation during the training. The Glorot uniform initializer~\cite{glorot10} and Adam optimizer~\cite{kingma14} with learning rate of 0.00001 is applied for training process. Once the training process is successful, the trained neural network can rapidly predict the 237600 test data samples. 
Fig.~\ref{fig:ml-prediction} compares the ML prediction with the GA solutions for all accumulated configurations, i.e. those used in the training phase and the remaining configurations used for validation. For all four local quantities, the NN gives rather good predictions as attested by the small MAE, which is of the order of less than one percent of the mean values for all quantities.

{\em Discussion and Outlook}. To summarize, we have introduced a ML model for predicting local electron correlation of Anderson-Hubbard Hamiltonian based on training a deep multi-task NN in configuration space. In order to describe the spatial inhomogeneity of the electronic structure, we use the real-space Gutzwiller method to numerically solve the AH model on a square lattice. Using the disorder potential in the neighborhood as the input, our ML trained NN is able to predict local electron properties such as double-occupancy and quasi-particle weight. Interesting phenomena such as the correlation induced screening of disorder potential and local Mott transition can be accurately predicted by our ML model. Our work provides a proof of principle study showing that deep NNs can serve as an efficient many-body problem solver for strongly correlated systems. For example, instead of the Gutzwiller solutions, one can train the NNs with data-sets obtained from the real-space DMFT or the VMC methods for the AH model. Although more computational effort is required to generate the training data, more accurate prediction can be achieved with the resultant NN model.

As discussed above, a primary motivation for ML trained NN is to bypass the expensive DFT calculation that is required in simulations such as {\em ab initio} molecular dynamics. Similarly, our proposed ML model as an efficient GA solver also has direct application for the molecular dynamics simulations of so-called Holstein-Hubbard model~\cite{holstein59,pradhan15,disante17}, in which the site potential $\varepsilon_i = - g X_i$ is related to the amplitude of local phonon mode $X_i$, here $g$ is the electron-phonon coupling constant. In such simulations~\cite{pradhan15}, forces acting on the local elastic modes are proportional to the local electron density $F_i = g \, n_i$, which can be efficiently computed using the trained NN. Another related application is to the recently proposed Gutzwiller molecular dynamics (GMD)~\cite{chern17}. The atomic forces in this method are computed from the optimized Gutzwiller many-electron wavefunction at every time step. Contrary to DFT-based molecular dynamics, GMD simulations allow one to investigate the effects of electron correlation on atomic structural dynamics~\cite{chern17}. Our work shows that ML techniques can be applied to develop a NN that efficiently emulates a GA solver. Preliminary results~\cite{maro18} indeed show that ML is a promising approach for such applications.  

\medskip

{\em Acknowledgment}. We thank Kipton Barros for useful discussions on ML methods. G.W.C. thanks Vladmir Dobrosavljevi\'c for insightful discussions on slave-boson and Gutzwiller methods for disordered Hubbard models. P.Z. and G.W.C. are partially supported by the Center for Materials Theory as a part of the Computational Materials Science (CMS) program, funded by the US Department of Energy, Office of Science, Basic Energy Sciences, Materials Sciences and Engineering Division. J.~Ma, Y. Tan and A.W.G. thank support from NSF-DMREF 1235230. The authors also acknowledge Advanced Research Computing Services at the University of Virginia for providing technical support that has contributed to the results in this paper.


\end{document}